\documentstyle[12pt]{article}

\def\a{\begin{eqnarray}}
\def\b{\end{eqnarray}}
\def\0{\nonumber}

\def\al{{\alpha}}
\def\lm{{\lambda}}

\renewcommand{\theequation}{\thesection.\arabic{equation}}

\def\omm{1--matrix model }

\def\mmm{multi--matrix models }

\def\dqg{2--dimensional quantum gravity }

\def\kph{{ KP hierarchy }}
\def\tauf{ the $\tau$--function }
\def\nls{{ non--linear Schr\"odinger hierarchy }}
\def\kdvh{{KdV hierarchy }}

\def\a{\begin{eqnarray}}
\def\b{\end{eqnarray}}
\def\0{\nonumber}
\def\ba{\begin{array}}
\def\ea{\end{array}}
\def\noal{\noalign{\vskip10pt}}

\def\al{{\alpha}}
\def\lm{{\lambda}}

\def\cf{{\cal F}}

\renewcommand{\theequation}{\thesection.\arabic{equation}}

\newlength{\extraspace}
\newlength{\extraspaces}
\newcounter{dummy}

\newcommand{\ai}{
\addtocounter{equation}{1}
\setcounter{dummy}{\value{equation}}
\setcounter{equation}{0}
\renewcommand{\theequation}{\thesection.\arabic{dummy}\alph{equation}}
\begin{eqnarray}
\addtolength{\abovedisplayskip}{\extraspaces}
\addtolength{\belowdisplayskip}{\extraspaces}
\addtolength{\abovedisplayshortskip}{\extraspace}
\addtolength{\belowdisplayshortskip}{\extraspace}}
\newcommand{\bj}{
\end{eqnarray}
\setcounter{equation}{\value{dummy}}
\renewcommand{\theequation}{\thesection.\arabic{equation}}}

\def\d{{\partial}}

\def\dtl{{\tilde\d}}
\def\lm{{\lambda}}
\def\dlm{{\partial \over {\partial \lm}}}

\def\tk{{\tilde K}}

\newcommand{\ddt}[1]{{\partial \over \partial t_{#1}}}
\newcommand{\ddy}[1]{{\partial \over \partial y_{#1}}}

\newcommand{\res}{{\rm res}\,}

\newcommand{\bac}{\begin{array}{c}}
\newcommand{\bacc}{\begin{array}{cc}}
\newcommand{\baccc}{\begin{array}{ccc}}
\newcommand{\barcl}{\begin{array}{rcl}}
\newcommand{\bacccc}{\begin{array}{cccc}}
\newcommand{\baccccc}{\begin{array}{ccccc}}
\newcommand{\baccccccc}{\begin{array}{ccccccc}}
\newcommand{\barclcrcl}{\begin{array}{rclcrcl}}
\newcommand{\bacl}{\begin{array}{cl}}
\newcommand{\bacll}{\begin{array}{cll}}
\newcommand{\eac}{\end{array}}

\newcounter{tabnum}
\begin{document}

\begin{flushright}
SISSA-ISAS 214/92/EP
\end{flushright}
\vskip0.5cm

\centerline{\LARGE The generalized KP hierarchy}
\vskip2.0cm
\centerline{\large C.S.Xiong\footnote{The address after December 1:
Inst. Theor. Phys., Academia Sinica, P.O.Box 2735, Beijing 100080, China}}
\vskip0.5cm
\centerline{International School for Advanced Studies (SISSA/ISAS)}
\centerline{Via Beirut 2, 34014 Trieste, Italy}
\vskip5cm
\abstract
{We propose one possible generalization of the  KP hierarchy, which
possesses multi bi--hamiltonian structures, and can be viewed as several KP
hierarchies coupled together.}

\vfill
\eject

\section{Introduction}

Non--linear integrable differential systems have been subject to intensive
investigations since last century. In the last few years,
one of the most exciting
developments in this field is the revelation of its close relationship
with 2--dimensional exactly sovable field theories, in particular
\dqg and string theory. This relation enables us to extract
non--perturbative properties of non--critical string. More precisely,
the discretization of \dqg can be reformulated as matrix models. Through
the {\it double scaling limit}, one can prove that \omm with even
potential is described by \kdvh and string equation.
Recently it has been shown that the double scaling limit is not an inevitable
step. In other words, the \kdvh is not merely an occasional effect of double
scaling limit but intrinsic in matrix model formulation.
The idea is as follows: we represent matrix models as certain
discrete linear
system(s), from which we can extract lattice integrable hierarchies, finally
we can directly extract differential hierarchies from these lattice
hierarchies. Using this approach, one can prove that
the \omm with general potential is characterized by \nls(NLS), which is
the two bosonic field representation of \kph.  In other words, the \omm gives
a new solution for \tauf of \kph\cite{BX1}.
In the same way, we can apply this procedure to \mmm to obtain their
full differential integrable hierarchies. In this process, we are led to
consider a possible generalization of \kph. The study of this generalization
is the main purpose of this letter. Actually another enlargement of \kph was
considered some years ago by adding additional flows. However, although
each of the additional flows commutes with all the old KP flows, they don't
commute among themselves\cite{Dickey}. Thus this generalization is not
consistent. Fortunately, we will see that properly introducing new flows
we can obtain an integrable hierarchy, in which all the flows are
commutative. We will explain the realization of such generalized \kph
in \mmm elsewhere\cite{BX2}.



\section{KP hierarchy}

Let us begin with a pseudo--differential operator(PDO) of arbitrary order
\a
A=\sum_{-\infty}^n a_i(x)\partial_i.\label{pseudodp}
\b
where $x$ is {\it the space coordinate}, while
$\d^{-1}$ is formal integral operation over $x$.
All the pseudo--differential operators form an algebra $\wp$ under the
generalized Leibnitz rules
\a
&&\d a(x)=a(x)\d+a'(x),\qquad [\d, x]=1,\0\\
&&\partial^{-1}\partial=\partial\partial^{-1}=1,\label{leibnitz}\\
&&\partial^{-j-1}a(x)=\sum_{l=0}^{\infty}(-1)^l{{j+l}\choose l}
a^{(l)}(x)\partial^{-j-l-1},\0
\b
where $a^{(l)}(x)$ denotes ${{\d^l a(x)}\over{\d x^l}}$.
The algebra $\wp$ has two sub--algebras:
\a
\wp=\wp_+\oplus\wp_-.\0
\b
where $\wp_+$ denotes the algebra of pure differential operators,
while $\wp_-$ means the algebra of pure integration operations.

For any given pseudo--differential operator $A$ of type (\ref{pseudodp}), we
call $a_{-1}(x)$ its residue, denoted by
\a
\res_{\d}A=a_{-1}(x)\qquad \hbox{ or}\qquad A_{(-1)}\0
\b
and we define the following functional
\a
<A>=\int a_{-1}(x)dx.\label{innerproduct}
\b
which naturally gives an inner scalar product on the algebra
$\wp$\cite{Babelon}.

\subsection{The integrable structure}

Now let $L$ be a pseudo--differential operator of the first order
\a
L=\d+\sum_{i=0}^{\infty}u_i(x)\d^{-i}\label{PDO}
\b
which we will call KP(or Lax) {\it operator}.
We call $u_i$'s KP {\it coordinates}.
$(L-\d)\in \wp_-$, so we can represent
 a functional of KP coordinates as
\a
f_X(L)=<LX>,\qquad\qquad X\in\wp_+\0
\b
\def\cf{${\cal F}(\wp_-)$ }
which  span a functional space \cf. The remarkable fact is that \cf is
invariant under the co--adjoint action of $\wp_+$, consequently the algebraic
structure on $\wp_+$ determines the Poisson structure on \cf
\a
\{f_X, f_Y\}_1(L)=L([X,Y])\label{poisson1}
\b
The infinite many conserved quantities (or Hamiltonians) are
\a
H_r={1\over r}<L^r>\qquad \forall r\geq1\label{hamiltonian}
\b
They generate infinite many flows,
\a
\ddt r L=[L^r_+, L]\label{KPequation}
\b
where the subindex ``+" indicates choosing the non--negative powers of $\d$.
Since
\a
[L^r_+, L]=[L, L^r_-]\in{\cal P}_-,\0
\b
we see that all the flows preserve the form of KP operator ``$L$", and
they all commute with each other. This commutativity implies
the ``{\it zero curvature representation}"
\a
\ddt m L^n_+-\ddt n L^m_+=[L^m_+, L^n_+],\qquad \forall n,m
\label{zerocurvature}
\b
By \kph we mean the set of differential equations (\ref{KPequation}) or
(\ref{zerocurvature}).
In fact the \kph possesses another Poisson structure\cite{Watanabe}
\a
\{f_X, f_Y\}_2(L)&=&<(XL)_+YL>-<(LX)_+LY>\0\\
&+&\int [L,Y]_{(-1)}\Bigl(\d^{-1}[L,X]_{(-1)}\Bigl)
\label{poisson2}
\b
With respect to these Poisson brackets,
the KP coordinates $u_i$ form $W$--infinity algebras.
The important point is that these two Poisson structures are compatible
in the sense
\a
\{f, H_{r+1}\}_1=\{f, H_r\}_2\qquad \forall\qquad\hbox{function f}
\b
This compatibility ensures the integrability of \kph. Generally speaking,
for a system of infinite many degrees of freedom, for example, the \kph,
we may  find various definitions of integrability\cite{Babelon}.
The essential point is that there must exist infinite many conserved
quantities in involution. Therefore we can list  some of the definitions
below
\begin{enumerate}
\item
{\it There exist two compatible Poisson brackets}
({\it or bi-hamiltonian structure}).
\item
{\it The flows are all commutative}.
\item
{\it There  exists the zero curvature representation}(\ref{zerocurvature}).
\end{enumerate}
For different purposes we may use different definitions. For example,
bi-Hamiltonian structure can exhibit the  Poisson algebraic structure of the
system. But in the next two sections we will mainly use the second definition
to prove the integrability of the generalized \kph due to its simplicity.

\subsection{The associated linear system}

The KP operator $L$ can be expressed in terms of the ``dressing" operator
\a
L=K\d K^{-1}\qquad K=1+\sum_{i=1}^{\infty}w_i\d^{-i}\0
\b
After defining
\a
\xi(t,\lambda)=\sum_{r=1}^{\infty}t_r\lambda^r
\b
and introducing the Baker--Akhiezer function
\a
\Psi(t,\lambda)=Ke^{\xi(t,\lambda)}\label{baker}
\b
we can associate to the KP hierarchy (\ref{KPequation}) a linear system
\a
\left\{\ba{l}
L\Psi=\lambda\Psi,\\\noal
\ddt r \Psi=L^r_+\Psi.
\ea\right.
\label{flow}
\b

\subsection{ The $\tau$--function}

One of the important ingrediants of the KP system is its $\tau$--function,
which can be introduced through Baker--Akhiezer function
\a
\Psi(t,\lambda)={{\tau(t_1-{1\over {\lambda}},t_2-{1\over
{2\lambda^2}},\ldots)}
\over{\tau(t)}}e^{\xi(t,\lambda)}
\b
One can prove that
\a
{\d^2\over{\d t_1\d t_r}}\ln\tau=\res_{\d}L^r,\qquad \forall r\geq1
\b
If we define a set of new functionals
\a
J_r=\int \res_{\d}L^r dx,\qquad \forall r\geq1
\b
then
\a
 \ddt s J_r=0,\qquad\quad \forall r,s\geq1
\b
So $J_r$'s are the conservation laws of the KP hierarchy.

\section{Generalization of KP hierarchy }

\setcounter{equation}{0}
\setcounter{subsection}{0}
\setcounter{footnote}{0}

Now we come to discuss the generalization of the \kph, which we promised
in the introduction.

\subsection{The additional flows}

Our purpose is to show
that we may introduce other series of flows. In order to do so, we define
a new operator\cite{Chenlee},
\a
M\equiv K(\sum_{i=1}^{\infty}rt_r\d^{r-1})K^{-1}=\sum_{i=-\infty}^{\infty}
v_i\d^i.\label{mpdo}
\b
which is conjugate to the KP operator $L$ in the sense that
\a
[L, M]=K [\d, \sum_{i=1}^{\infty}rt_r\d^{r-1}]K^{-1}=1,\label{conjugate}
\b
We can derive the equations of motion for $M$,
\a
\left\{\begin{array}{l}
\ddt r M=[L^r_+, M],\\\noalign{\vskip10pt}
\dlm\Psi=M\Psi.
\end{array}\right.
\b
So we see that $L$ and $M$ are nothing but the operatorial expressions of
$\lm,\dlm$ (acting on  $\Psi(t,\lm)$).

As we know, the basic requirement for new flows is that they should preserve
the form of KP operator $L$. So we can define new flows like
\a
\ddt {mn} L=[L, (M^mL^n)_-],\qquad\forall m,n.\label{additional}
\b
One can show that each flow commutes with KP flows (\ref{KPequation}), but
these additional flows do not commute among themselves\cite{Dickey}. Our aim
is to show that properly choosing combinations of these additional flows, we
can define new flows which commute with the old KP flows and among themselves.

\subsection{Another series of flows}

Our  starting  remark is that the $t$--series of perturbations is due to
the fact that $[L, L^r_-]\in{\cal P}_-,\forall r\geq1$. Now
we also have $[L, M^r_-]\in{\cal P}_-,\forall r\geq1$,
so we could introduce a new series of deformation
parameters\footnote{That is to say, the KP coordinates $u_i$'s and $v_i$'s
depend on both $t$ and $y$.} $y_1,y_2,y_3,\ldots$, such that
\a
\left\{\begin{array}{l}
\ddy r L=[L, M^r_-],\\\noalign{\vskip10pt}
\ddy r \Psi=-M^r_-\Psi.
\end{array}\right.
\b
All of these equations together result in the following enlarged KP system
\a
\left\{\begin{array}{l}
\ddt r L=[L^r_+, L],\\\noalign{\vskip10pt}
\ddt r M=[L^r_+, M],\\\noalign{\vskip10pt}
\ddy r L=[L, M^r_-],\\\noalign{\vskip10pt}
\ddy r M=[M^r_+, M].
\end{array}\right.\label{gen}
\b

Now we  should prove  that these new series of perturbations do not distroy
consistency, that is to say,
we should check the commutativity of all these flows. In the following we only
consider an example, i.e.
\a
\ddt r\bigl(\ddy s L\bigl)=\ddy s\bigl(\ddt r L\bigl).\label{comflows}
\b
Using eqs.(\ref{gen}), we see that the left hand side is
\a
&&\quad \ddt r\bigl(\ddy s L\bigl)=\ddt r[L, M^s_-]\0\\
&&=[[L^r_+, L], M^s_-]+[L,[L^r_+, M^s]_-]\0\\
&&=[[L^r_+, L], M^s_-]+[L,[L^r_+, M^s_-]]-[L,[L^r_+, M^s_-]_+]\0\\
&&=[L^r_+, [L, M^s_-]]+[[L^r,M^s_-]_+, L]\0\\
&&={\rm r.h.s.}\0
\b
The other cases can be checked in the similar way.
So the perturbations we introduced before indeed give an enlarged \kph.
Its associated linear system is
\a
\left\{\begin{array}{l}
L\Psi=\lm\Psi,\\\noalign{\vskip10pt}
\ddt r \Psi=L^r_+\Psi,\\\noalign{\vskip10pt}
\ddy r \Psi=-M^r_-\Psi,\\\noalign{\vskip10pt}
M\Psi=\dlm\Psi.
\end{array}\right.\label{lmpsi}
\b

The usual KP hierarchy (\ref{KPequation}) is a particular case
of eqs.(\ref{gen}) by fixing the $y$--series of the perturbations.

\subsection{The new basic derivative and the new bi--hamiltonian structure}

As we remarked a moment ago, when we disgard the $y$--series of flows, we
recover the usual \kph, whose hamiltonians are
\a
H_{r(L)}={1\over r}<L^r>,\0
\b
here we use the subindex $(L)$ to indicate that the Hamiltonians are
constructed
from the KP operator $L$. We may  also use the same symbol to denote the
Poisson
brackets, $\{, \}_{(L)}$.

Now if we fix all the $t$--series of parameters,  then we get another subset of
the enlarged hierarchy (\ref{gen}), that is
\a
\left\{\begin{array}{l}
\ddy r L=[L, M^r_-],\\\noalign{\vskip10pt}
\ddy r M=[M^r_+, M].
\end{array}\right.\label{geny}
\b
The second equation is in fact a \kph with KP operator $M$ of the form
(\ref{mpdo}). Since all these flows commute, this is an integrable
system, and there should  exist two compatible Poisson brackets written
in terms of coordinates $v_i$'s. However, this bi--hamiltonian structure
is unknown due to the  fact that the positive powers of $\d$ in $M$ go to
infinity.


Fortunately, we may overcome the difficulty by introducing a new basic
derivative. To this end, we recall that in our
previous analysis we treated $t_1$ as the space coordinate.
For later convenience, we denote $\ddy 1$ by $\dtl$.
{}From the $y_1$--flows of $\Psi$, we may extract an operator identity
\a
\dtl=-M_-=\sum_{i=1}^{\infty}\Gamma_i\d^{-i}.\label{dtilde}
\b
Since any positive powers of $\dtl$ belongs
to ${\cal P}_-(\d)$, so $\{\dtl^i;i\geq1\}$
forms a basis of ${\cal P}_-(\d)$.
We may invert the relation (\ref{dtilde}) to express $\d$ in
terms of the new derivative\footnote{
Rigorously speaking, this is only true when it acts on the function $\Psi$.
But we  may think of it in the following way, starting from
\a
\dtl\Psi=\sum_{i=1}^{\infty}\Gamma_i\d^{-i}\Psi.\0
\b
properly choosing the combinations of $\dtl$ such that we can
reexpress the $\d^{-1}\Psi$ in terms of new derivatives
$\dtl$, we replace  all the derivatives $\d$ in the linear system
(\ref{lmpsi}) by $\dtl$. So we may interpret $y_1$ as another space
coordinate.} $\dtl$
\a
\d=\sum_{i=1}^{\infty}{\tilde \Gamma}_i\dtl^{-i}.\label{d}
\b
Using this fact, we get
\a
M=-(\dtl+\sum_{i=1}^{\infty}{\tilde v}_i\dtl^{-i})=-\tk\dtl\tk^{-1}.
\b
with new dressing operator $\tk$ and new KP coordinates ${\tilde v}_i$'s.
Obviously
\a
M^r_-(\d)=M^r_+(\dtl),\qquad \forall r\geq1,
\b
where LHS is expanded in powers of $\d$, while the RHS is expanded in
powers of $\dtl$.  Using eq.(\ref{d}),
we can reexpress all the formulas (\ref{gen}) in terms of this
new derivative $\dtl$, i.e.
\a
\left\{\begin{array}{l}
\ddy r (-M(\dtl))=(-1)^{r+1}[(-M)^r_+(\dtl),
(-M)(\dtl)],\\\noalign{\vskip10pt}
\ddy r L(\dtl)=(-1)^{r+1}[(-M)^r_+(\dtl), L(\dtl)],\\\noalign{\vskip10pt}
\ddt r (-M)(\dtl)=-[(-M)(\dtl), L^r_-(\dtl)],\\\noalign{\vskip10pt}
\ddt r L(\dtl)=[L^r_+(\dtl), L(\dtl)].
\end{array}\right.
\b
Apart from some additional signs, these equations are isomorphic to
eqs.(\ref{gen}).
This reminds us that we can even consider $(-M)$ as a KP operator, and
alternatively interpret $y_1$ as space coordinate, all the other parameters
as time parameters.
Therefore we can define two compatible Poisson brackets
for KP operator $(-M)$ by simply replacing $L$ in (\ref{poisson1}) and
(\ref{poisson2}) by $(-M)$, which shows that
on the space $y_1$, the fields ${\tilde v}_i$'s form
$W_{\infty}$ algebras too.

\section{Further perturbations and the full generalized \kph}

\setcounter{equation}{0}
\setcounter{subsection}{0}
\setcounter{footnote}{0}

In the previous section we have shown that the \kph can be perturbed by
the conjugate operator $M$ of the KP operator $L$.
In fact, the KP system allows further deformations.

\subsection{The new series of the flows}

In order to explain the further perturbations just mentioned,
we change a little bit our notation. Denote $t_r$'s and $y_r$'s  by
$t_{1r}$ and $t_{2r}$ respectively. Furthermore define
\a
&&L(1)\equiv L,\qquad\qquad V(1)\equiv\sum_{r=1}^{\infty}rt_{1r}L^{r-1}(1)\0\\
&&L(2)\equiv -{1\over {c_{12}}}M\qquad V(2)\equiv\sum_{r=1}^{\infty}
rt_{2r}L^{r-1}(2)\0
\b
Now let us introduce new operators in the following way
\ai
&&L(\alpha)\equiv-{1\over c_{\alpha-1,\alpha}}\Bigl(c_{\alpha-2,\alpha-1}
L(\alpha-2)+V(\alpha-1)\Bigl)\label{loperator}\\
&&V(\alpha)=\sum_{r=1}^{\infty}rt_{\alpha,r}L^{r-1}(\alpha),\qquad \alpha=3,4,
\ldots,n
\bj
where $c_{\al,\al+1}$'s are arbitrary constants, which amount to rescaling
 the space coordinates, and $n$ is an arbitrary positive integer.
Then, in the same way, we can perturb the system further as follows
\ai
&&\ddt {\beta r} L(\alpha)=[L^r_+(\beta), L(\alpha)],\qquad 1\leq\beta<\alpha
\label{flowa}\\
&&\ddt {\beta r} L(\alpha)=[L(\alpha), L^r_-(\beta)],\qquad
\alpha\leq\beta\leq n\label{flowb}
\bj
Now in order to justify the consistency of these perturbations, we once again
should prove that all the flows commute among themselves. Let us check one
example,
\a
\ddt{\al l}\bigl(\ddt{\beta m}L(\gamma)\bigl)
=\ddt{\beta m}\bigl(\ddt{\al l}L(\gamma)\bigl),\qquad\al<\beta<\gamma.\0
\b
Using the above hierarchy and Jacobi identities, we see that
\a
{\rm l.h.s.}=\ddt{\al l}[L^m_+(\beta), L(\gamma)]
=[[L^l_+(\al), L^m(\beta)], L(\gamma)]+[L^m_+(\beta), [L^l_+(\al),
L(\gamma)]].\0
\b
and
\a
{\rm r.h.s.}=\ddt{\beta m}[L^l_+(\al), L(\gamma)]
=[[L^l(\al), L^m_-(\beta)]_+, L(\gamma)]+[L^l_+(\al),
[L^m_+(\beta), L(\gamma)]].\0
\b
The first term of ``l.h.s" can be written as
\a
{\rm the~~1st~~term}&=&[[L^l_+(\al), L^m_+(\beta)], L(\gamma)]
+[[L^l_+(\al), L^m_-(\beta)]_+, L(\gamma)]\0\\
&=&[[L^l_+(\al), L^m_+(\beta)], L(\gamma)]+
[[L^l(\al), L^m_-(\beta)]_+, L(\gamma)],\0
\b
therefore
\a
{\rm l.h.s.}
&=&[[L^l(\al), L^m_-(\beta)]_+, L(\gamma)]
+[[L^l_+(\al), L^m_+(\beta)], L(\gamma)]+
[L^m_+(\beta), [L^l_+(\al), L(\gamma)]]\0\\
&=&[[L^l(\al), L^m_-(\beta)]_+, L(\gamma)]
+[L^l_+(\al), [L^m_+(\beta), L(\gamma)]]\0\\
&=&{\rm r.h.s.}\0
\b
All the other cases can be done in the same way. Therefore eqs.(4.2)
really define an integrable system.
The associated linear system is
\a
\left\{\begin{array}{l}
L(1)\Psi=\lm\Psi,\\\noalign{\vskip10pt}
\ddt {1,r}\Psi=L^r_+(1)\Psi,\\\noalign{\vskip10pt}
\ddt {\al,r}\Psi=-L^r_-(\al)\Psi,\qquad \al=2,3,\ldots,n,\\\noalign{\vskip10pt}
M\Psi=\dlm\Psi.
\end{array}\right.\label{lmpsigen}
\b
In fact we can rewrite this
linear system in a better way by choosing a new function
\a
\Psi(\lm,t)\Longrightarrow\xi(\lm,t)=
\exp(-\sum_{r=1}^{\infty}t_{1,r}\lm^r_1)\Psi(\lm,t),\0
\b
then all the flows can be
summarized by a single equation
\a
\ddt {\al, r}\xi=-L^r_-(\al)\xi.
\b
The consistency conditions of the above linear system exactly give the
hierarchy (4.2) We would like to remark
that the hierarchy (4.2) have several important
sub--hierarchies.

$(i)$. $\al=\beta=1$, the eqs.(\ref{flowb}) are nothing but the
usual \kph (\ref{KPequation}).

$(ii)$. $2\leq\al=\beta\leq  n$, the eqs.(\ref{flowb}) give $(n-1)$
KP hierarchies whose KP operator possess the form (\ref{loperator}).

$(iii)$. All the  flows commute.

$(iv)$. When $n\longrightarrow\infty$, the full hiearchy (4.2)
contains all possible combinations of additional flows
(\ref{additional}).

We may  conclude that  the hierarchy (4.2)
possess $n$ bi--hamiltonian structures, each  of  them generates a \kph,
all of these hierarchies couple together. The integrability of the system
is guaranteed by the commutativity of the flows. However, we are not sure
whether the  hamiltonians in
different series are commutative.

\subsection{New bi--hamiltonian structures}

In the above analysis, all the operators are expanded in terms of $\d$.
However,
if we  use the same trick as the one in previous section,
it is not difficult to reexpress them in terms of any one of $\ddt {\al,1}$'s.
Let us define
\a
\d_{\al}\equiv {\d\over{\d t_{\al,1}}}
\b
and expand $L_-(\al)$ in powers of $\d$
\a
L_-(\al)=-\sum_{i=1}^{\infty}\Gamma^{(\al)}_i\d^{-i}
\b
then the first flows of the linear system (\ref{lmpsigen}) suggest
\a
\d_{\al}=\sum_{i=1}^{\infty}\Gamma^{(\al)}_i\d^{-i}
\b
similar to the argument in the previous section, we can invert these
relations, such that
\a
\d=\sum_{i=1}^{\infty}{\tilde\Gamma}^{(\al)}_i\d^{-i}_{\al}
\b
Substituting them into the formulas (\ref{loperator}), we get the expansions
of $L(\al)$ in $\ddt {\beta,1}$ ({\it for any $\al,\beta$}). In particular
$L(\al)$ expanded in $\ddt {\al,1}$ is also a KP operator,
\a
L(\al)=-(\d_{\al}+\sum_{i=1}^{\infty}v^{(\al)}_i\d^{-i}_{\al})
\b
and its $\al-th$ series of flows is nothing but the ordinary KP hierarchy
\a
\ddt {\al, r} L(\alpha)=(-1)^{r+1}[L^r_+(\alpha), L(\al)]
\b
where the operators are expanded in powers of $\d_{\al}$, and the additional
sign indicates rescaling of the parameters. Of course,
for this subsystem, we can
construct its integrable structure, by replacing $L$ in (\ref{poisson1})
and (\ref{poisson2}) by $L(\al)$.
Therefore, we may say
that KP system (4.2) possesses multi bi--hamiltonian structures,
and it contains ``$n$" coupled ordinary KP hierarchies.
The coupling comes from the dynamical equations (4.2)
with $\al\neq\beta$.

\subsection{The $\tau$--function of the generalized \kph}

Using eqs.(4.2), we get
\a
\ddt {\beta,s} \res_{\d}L^r(\al)
=\ddt {\al,r} \res_{\d}L^s(\beta),\qquad \forall \al,\beta;\quad r,s.
\b
These equalities imply the existence of $\tau$--function
\a
{\d^2\over{\d t_{1,1}\d t_{\al,r}}}\ln\tau=\res_{\d}L^r(\al),\qquad
\forall \al,r.\label{taumulti}
\b
Using this $\tau$--function, we can introduce a series of the Baker--Akhiezer
functions,
\a
\Psi_{\al}(t,\lm_{\al})={{\tau(t_{\al,1}-{1\over {\lm_{\al}}},t_{\al,2}-
{1\over {2\lm^2_{\al}}},
\ldots)}\over{\tau(t)}}e^{\xi(t,\lm_{\al})}
\b
where $\al=1,2,\ldots, n$. To each $\Psi_{\al}$ we can associate a linear
system. Among them, the $\al=1$ case was discribed above. The other
cases can be analysed in the similar way\cite{Thesis}.

\section{Discussion}

We have shown that the KP hierarchy can be extended to a much larger hierarchy
by introducing additional KP operators. This generalized hierarchy can be
considered as several coupled KP hierarchies. For each
of the KP operators, we have constructed its  bi--hamiltonian structure
by introducing new basic derivatives. Although we do not
know if all these hamiltonians are in involution,
The commutativity of the flows guarantees the integrability of the system.

As we know in the ordinary KP hierarchy case, the series of flows reflects the
large symmetry of the system generated  by its Hamiltonians. In our case,
the multi--series of flows imply that this new hierarchy (4.2)
should possess a much larger symmetry.
However we are not sure what this large symmetry is.

It is not clear if this new hierarchy relates to the multi--component
KP hierarchy. Another interesting problem is to reduce this hierarchy to the
known hierarchies like generalized KdV hierarchies. This is under
investigation.

\vskip0.8cm
\noindent
{\bf Achnowledgement}

\vskip 0.2cm

I would like to thank Prof. L. Bonora for his constant encouragement,
valuable suggestions and fruitful discussions.

\renewcommand{\Large}{\normalsize} 
\end{document}